\documentclass[%
 reprint,
 amsmath,amssymb,
 aps,
]{revtex4-2}

\usepackage{graphicx}
\usepackage{dcolumn}
\usepackage{bm}
\usepackage{lipsum}
\usepackage{color}

\usepackage[usenames,dvipsnames]{xcolor}
\usepackage[colorlinks, linkcolor=blue!50!black, urlcolor=blue!50!black, citecolor=blue!50!black]{hyperref}

\newcommand{\MO}[1]{\textcolor{red}{#1}}

\begin{document}

\setlength{\belowdisplayskip}{3pt} \setlength{\belowdisplayshortskip}{3pt}
\setlength{\abovedisplayskip}{3pt} \setlength{\abovedisplayshortskip}{3pt}

\title{Multiscale Entropies as Order Parameters for Nonequilibrium Phase Transitions}

\author{Gerhard Jung}
\email[Corresponding author: ]{gerhard.jung.physics@gmail.com}
\affiliation{Institut f\"ur Theoretische Physik, Universit\"at Innsbruck, 6020 Innsbruck, Austria}

\author{Misaki Ozawa}
\affiliation{Univ. Grenoble Alpes, CNRS, LIPhy, 38000 Grenoble, France}

\date{\today}

\begin{abstract}
We develop a multiscale entropy framework based on the wavelet conditional renormalization group that decomposes the total Shannon entropy into contributions from different spatial scales. Generative modeling enables accurate estimation of these entropies from configurations or snapshots, even in nonequilibrium systems whose underlying probability distributions are unknown. Applications to the equilibrium $\varphi^4$ model and nonequilibrium active Model B+ show that multiscale entropies reveal criticality and spatial organization obscured in the total entropy. These entropies serve as scale-dependent order parameters for characterizing phase transitions and pattern formation.

\end{abstract}

\maketitle


{\it Introduction}: Entropy quantifies the volume of accessible states in phase space and thus provides a useful thermodynamic measure to characterize many physical systems. For example, the emergence of order through pattern formation or collective behavior is generally accompanied by a change in entropy. Moreover, nonanalytic behavior of the entropy or its derivatives, such as a discontinuity or a kink, signals the presence of a phase transition~\cite{callen1993thermodynamics,chaikin1995principles,sethna2021statistical}.

In experiments, entropy can be measured indirectly, for example by integrating the specific heat~\cite{takahara1995calorimetric}. 
In computer simulations, it is commonly estimated by thermodynamic integration, using sampling algorithms that assume thermal equilibrium~\cite{frenkel2023understanding}. 
These approaches are powerful but rely on the knowledge of an equilibrium thermodynamic structure.

Recently, data-compression algorithms have been proposed as an alternative route to estimate entropy directly from configurations or snapshots~\cite{martiniani2019quantifying,avinery2019universal,zu2020information,martiniani2020correlation,cavagna2021vicsek,douglass2024complexity,fraenkel2024information,javerzat2026entropy,diamant2026perspective}.
In these methods, the Shannon entropy, which coincides with the thermodynamic entropy in thermal equilibrium, is estimated using efficient lossless compression algorithms.
This strategy is particularly appealing because it can be applied both to systems for which no appropriate order parameter is known \textit{a priori} and to nonequilibrium systems, for which the underlying probability distribution is generally unknown.
These studies have successfully estimated the total entropy of nonequilibrium systems and shown that its evolution with a control parameter can provide signatures of phase transitions.
Furthermore, combining data compression with decimation or spatial subsampling makes it possible to probe spatial correlations and order within the system in a large variety of different systems~\cite{feldman2003structural,cammarota2012patch,dunleavy2012using,gleiser2013information,martiniani2020correlation}. Entropy-estimation methods based on recurring spatial patterns, requiring only particle coordinates as input, have also been proposed~\cite{javerzat2026entropy}.

Another route to estimating entropy, or equivalently free-energy-related quantities, in both equilibrium and nonequilibrium systems is provided by machine learning~\cite{wu2019solving,nir2020machine,nicoli2021estimation,gu2022thermodynamics,gelman2024nonequilibrium}.
In particular, generative models provide a data-driven means of approximating an underlying probability distribution, which is generally unknown, from samples. Once trained, these models can be used to estimate the entropy of the learned distribution~\cite{gu2022thermodynamics,gelman2024nonequilibrium}.

Although these generative-modeling approaches are promising and can successfully estimate entropy, they do not by themselves provide direct information about the spatial scales associated with ordering and pattern formation.
This is a significant limitation because spatial organization is often a key aspect of phase transitions in physical systems~\cite{chaikin1995principles,sethna2021statistical}, as exemplified by diverging correlation lengths near critical points and by the emergence of patterns and domains over multiple length scales in nonequilibrium systems~\cite{cross1993pattern,onuki2002phase,tjhung2018cluster}.

In this paper, we develop a set of multiscale entropies based on generative modeling within the framework of the wavelet conditional renormalization group (WCRG)~\cite{marchand2023multiscale,guth2023conditionally,brossollet2025effective,lempereur2026hierarchic,bandini2026overcoming}.
The central idea is to compute the Shannon entropy by expressing the underlying probability distribution as a product of conditional probability distributions associated with different length scales.
Accurate estimates of these conditional distributions, and hence of the associated entropies, are obtained by learning, scale by scale, the conditional energy function (or Hamiltonian).
This hierarchical representation not only provides an accurate estimate of the total entropy but also naturally introduces two complementary scale-dependent quantities: the multiscale entropy and the conditional entropy. Together, they reveal how different length scales contribute to the total entropy, providing physical insight into spatial ordering phenomena, including nonequilibrium phase transitions. They are particularly useful when no appropriate order parameter is known \textit{a priori}, as they can characterize spatial structures across multiple length scales, in a similar spirit as the data-compression-based approach of Ref.~\cite{martiniani2020correlation}.

We apply this method to the equilibrium $\varphi^4$ model~\cite{milchev1986finite,troster2005free}, which exhibits a second-order phase transition, and to the active Model B+~\cite{tjhung2018cluster}, which exhibits nonequilibrium phase transitions.
Our method successfully estimates the multiscale entropies and characterizes phase transitions associated with spatial ordering in both equilibrium and nonequilibrium systems.

{\it Methods}: We first briefly review the wavelet conditional renormalization group (WCRG) developed in Ref.~\cite{marchand2023multiscale}. We consider a field $\varphi_0 \in \mathbb{R}^{L^2}$ defined on a two-dimensional square lattice of linear size $L$, with periodic boundary conditions. 
The discrete wavelet transform (see Appendix~\ref{sec:DWT}) provides a multiscale representation of the field by coarse-graining it over $4^j$ lattice sites. We therefore refer to $j$ as the scale index. At each scale, it decomposes the field into a coarse-grained field $\varphi_j \in \mathbb{R}^{L_j^2}$, defined on a two-dimensional square lattice of linear size $L_j = L/2^j$
and a wavelet field $\overline{\varphi}_j \in \mathbb{R}^{3L_j^2}$ containing the fine-scale fluctuations.
At each scale $j$, the coarse-grained field $\varphi_j$ is obtained from $\varphi_{j-1}$ by applying a low-pass filter, while the wavelet field $\overline{\varphi}_j$ is obtained by applying a high-pass filter. Conversely, $\varphi_{j-1}$ can be reconstructed from $\varphi_j$ and $\overline{\varphi}_j$ by applying the corresponding transposed filters. The filters can be chosen such that the coarse-grained and wavelet components form an orthogonal decomposition. Thus, the two representations, $\varphi_{j-1}$ and $(\varphi_j,\overline{\varphi}_j)$, are equivalent and can be regarded as related by an invertible change of variables (see Fig. 2 of Ref.~\cite{marchand2023multiscale} for visual illustration).

With this representation, the probability distribution of the microscopic field,
$p_0(\varphi_0)$, can be factorized into conditional probability distributions
$\overline{p}_j(\overline{\varphi}_j|\varphi_j)$ across scales $j=1,2,\ldots,J$~\cite{marchand2023multiscale} (see Appendix~\ref{sec:multiscale}):
\begin{equation}
p_0(\varphi_0)
=
\alpha \, p_J(\varphi_J)
\prod_{j=1}^J
\overline{p}_j(\overline{\varphi}_j|\varphi_j),
\end{equation}
where $p_j(\varphi_j)$ denotes the probability distribution of the
coarse-grained field $\varphi_j$, and $\alpha$ is a normalization factor
associated with the change of variables.
We consider $J$ to be the most coarse-grained scale such that only a single site remains, i.e., $L_J=L/2^J=1$.

This factorization of the microscopic probability distribution naturally leads to our central proposal: a decomposition of the total Shannon entropy into contributions from different scales (see Appendices~\ref{sec:energy_based} and~\ref{sec:multiscale_entropy}):
\begin{align}
    S_0
    &= - \int d\varphi_0 \, p_0(\varphi_0)\ln p_0(\varphi_0) + \ln a_0 \\
    &= S_J + \sum_{j=1}^J \overline{S}_j ,\label{eq:S_0_WCRG}
\end{align}
where $S_j$ and $\overline{S}_j$ denote the \textit{multiscale entropy} and \textit{wavelet conditional entropy} at scale $j$, respectively:
\begin{align}
S_j
&=
-\int d\varphi_j \,
p_j(\varphi_j)
\ln p_j(\varphi_j) + \ln a_j  ,
\label{eq:S_j}
\\
\overline{S}_j
&=
-\hspace{-0.1cm}\int \hspace{-0.1cm} d\varphi_j 
p_j(\varphi_j)
\hspace{-0.1cm} \int \hspace{-0.1cm} d\overline{\varphi}_j 
\overline{p}_j(\overline{\varphi}_j|\varphi_j)
\ln \overline{p}_j(\overline{\varphi}_j|\varphi_j) + \ln b_j.
\end{align}
Here, $a_j$ and $b_j$ are known constants originating from normalization factors at each scale. 
The multiscale and conditional entropies are related through
\begin{equation}
\overline{S}_j
=
S_{j-1} - S_j,
\label{eq:entropy_recurrence_main}
\end{equation}
such that $\overline{S}_j$ quantifies the entropy difference between two successive scales. The numerical methods used to estimate $\overline{S}_j$ and $S_j$ at different scales are described in the Appendix~\ref{sec:multiscale_entropy}.

We note that the term ``multiscale entropy'' has also been widely
used in time-series analysis, where sample entropy is evaluated for
a sequence of temporally coarse-grained signals
\cite{costa2002multiscale,humeau2015multiscale}. This approach has
also been extended to two-dimensional image analysis~\cite{silva2018two}.
Although similar in spirit, these approaches should be distinguished
from the present framework, which considers the Shannon entropy of
spatial probability distributions and decomposes it into contributions
associated with different length scales.

\begin{figure}
\includegraphics[width=\linewidth]{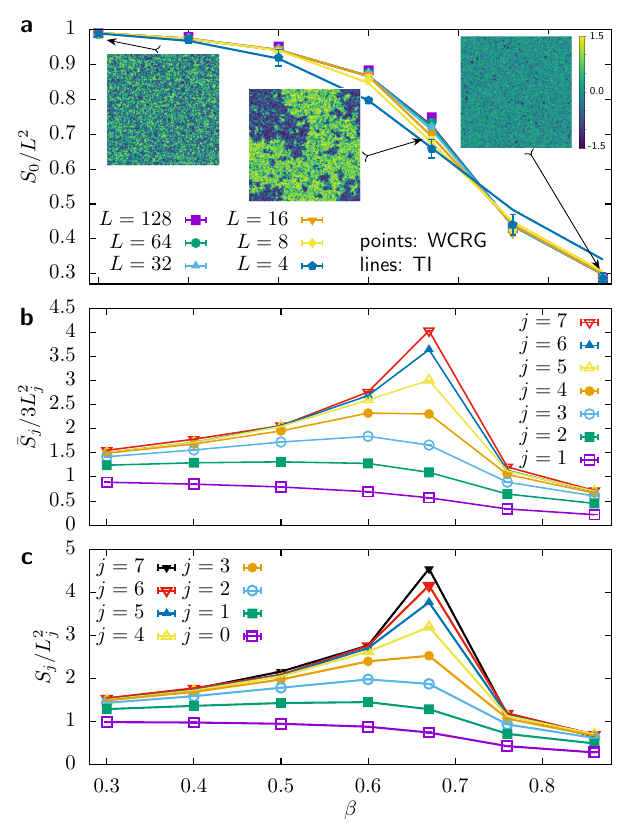}
\caption{Multiscale and conditional entropies per site in the $\varphi^4$ model for different values of the control parameter $\beta$.
(a) Comparison of the total entropy $S_0$ obtained by thermodynamic integration (TI) of the microscopic model and by the wavelet conditional renormalization group (WCRG) model for different system sizes $L=4,8,16,32,64,128$. Snapshots show representative configurations at $\beta=0.3$, $0.67$, and $0.86$ (same color code) for $L=128$.
(b) Conditional entropies $\overline{S}_j$ extracted from the WCRG model at scales ranging from $j=1$ to $j=J=7$ for $L=128$.
(c) Multiscale entropies $S_j$ for the same system as in panel~(b).}
\label{fig:phi4_entropy}
\end{figure}

{\it $\varphi^4$ model}: We first study the lattice $\varphi^4$ field model~\cite{milchev1986finite,troster2005free}, which is a representative equilibrium system for investigating critical phenomena. The model is parametrized by a control parameter $\beta$, which plays a role analogous to the inverse temperature~\cite{kaupuvzs2016corrections}: small $\beta$ corresponds to the disordered phase, whereas large $\beta$ corresponds to the ordered, symmetry-broken phase. The system undergoes a continuous phase transition at $\beta_c \approx 0.67$, where the correlation length diverges (see snapshots in Fig.~\ref{fig:phi4_entropy}(a)). The dataset used for the WCRG modeling of the $\varphi^4$ model consists of many independent configurations generated by Monte Carlo simulations. Details of the numerical implementation of the WCRG learning procedure, the entropy extraction and of the $\varphi^4$ model are provided in Appendices~\ref{app:learning_WCRG}, \ref{app:extracting_entropy},  and~\ref{app:phi4_model}, respectively.

Figure~\ref{fig:phi4_entropy}(a) shows the total entropy per lattice site, $S_0/L^2$, for several system sizes $L$. We compare the entropy estimated using the WCRG generative model, Eq.~(\ref{eq:S_0_WCRG}), with an independent baseline estimate obtained by thermodynamic integration using equilibrium Monte Carlo simulations of the microscopic model~\cite{milchev1986finite,troster2005free}. The entropy $S_0/L^2$ decreases monotonically with increasing $\beta$, reflecting the progressive ordering of the configurations. We find that the WCRG estimates closely agree with the thermodynamic-integration results obtained from the microscopic model, thereby validating our method. 

The advantage of the hierarchical entropy decomposition in Eq.~(\ref{eq:S_0_WCRG}) is that it resolves the contributions to the total entropy from different length scales. Figure~\ref{fig:phi4_entropy}(b) shows the conditional entropy per site, $\overline{S}_j/(3L_j^2)$, which quantifies the contribution associated with each scale $j$. At small $j$, $\overline{S}_j/(3L_j^2)$ decreases monotonically with increasing $\beta$, similarly to the total entropy $S_0/L^2$. As $j$ increases, however, a pronounced peak develops near the critical point $\beta_c$. This is particularly useful because the critical behavior is difficult to identify directly from the smooth variation of the total entropy $S_0/L^2$. In contrast, $\overline{S}_j/(3L_j^2)$ at larger scales echos the critical behavior, demonstrating that the hierarchical decomposition encodes information about criticality that is not readily visible in the total entropy alone.

We also monitor the multiscale entropy per site, $S_j/L_j^2$, in Fig.~\ref{fig:phi4_entropy}(c), where $j=0$ corresponds to the total entropy $S_0/L^2$ shown in Fig.~\ref{fig:phi4_entropy}(a). As the scale $j$ increases, $S_j/L_j^2$ develops a pronounced nonmonotonic behavior, with a peak emerging around $\beta_c$ and becoming more pronounced at larger scales. This behavior signals the critical phenomenon, similarly to $\overline{S}_j/(3L_j^2)$. According to Eq.~(\ref{eq:entropy_recurrence_main}), the difference in multiscale entropy between two successive scales gives rise to the conditional entropy $\overline{S}_j$.

The nonmonotonic behavior of the multiscale entropy, particularly at the largest scale $S_J$ shown in Fig.~\ref{fig:phi4_entropy}(c), is reminiscent of that of the susceptibility, $\chi=L^2\left(\langle m^2\rangle-\langle m\rangle^2\right)$,
where
$m=L^{-2}|\sum_i \varphi_0(i)|$ is the magnetization.
We now present a phenomenological argument suggesting that $S_J$ encodes essentially the same information as the susceptibility $\chi$ and the correlation length $\xi$, which are related by $\chi \sim \xi^{2-\eta}$.
At the largest scale, $\varphi_J$ is proportional to the average of the microscopic field variables, and hence to the global order parameter $m$: $\varphi_J=\Gamma^{-1}\sqrt{L^2}\,m$,
where $\Gamma=\prod_{j=1}^J \gamma_j$ is a normalization constant.
Thus, it follows that ${\rm Var}[\varphi_J]=\Gamma^{-2}\chi$. If the coarse-grained distribution $p_J(\varphi_J)$ is well approximated by a Gaussian, as confirmed numerically~\cite{marchand2023multiscale}, its entropy in Eq.~(\ref{eq:S_j}) is determined solely by the variance ${\rm Var}[\varphi_J]$. We then obtain
\begin{equation}
S_J
\approx
\frac{1}{2}\ln\chi
+
\frac{1}{2}\ln\left(\frac{2\pi e}{\Gamma^2}\right) + \ln a_J.
\label{eq:SJ_chi}
\end{equation}
Using the scaling relation $\chi \sim \xi^{2-\eta}$ for the correlation length $\xi$, we also obtain $S_J \approx \frac{2-\eta}{2}\ln \xi + \mathrm{const}$.

\begin{figure}
\includegraphics[width=\linewidth]{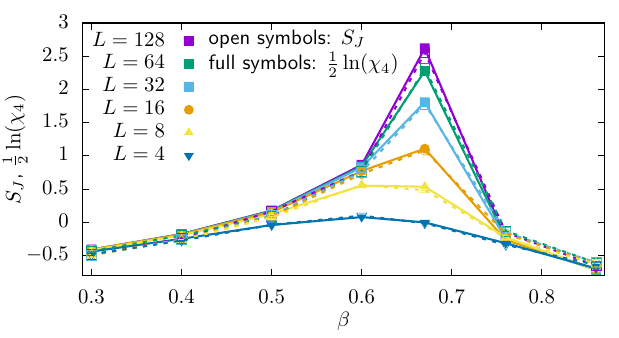}
\caption{Coarse-grained entropy $S_J$ and susceptibility $\chi_4$ for the $\varphi^4$ model for different control parameters $\beta$ and different system sizes $L.$ $S_J$ has been shifted to account for different normalization. Lines are a guide for the eye.}
\label{fig:susceptibility}
\end{figure}

To test this argument, we compare $S_J$ and $\frac{1}{2} \ln\chi$ in Fig.~\ref{fig:susceptibility}.
We find remarkable agreement between them up to an additive constant, confirming that the growth of $S_J$ captures the enhancement of the susceptibility. This demonstrates that our multiscale-entropy approach can characterize critical behavior even in situations where no appropriate order parameter is known \textit{a priori}.

\begin{figure}
\includegraphics[width=\linewidth]{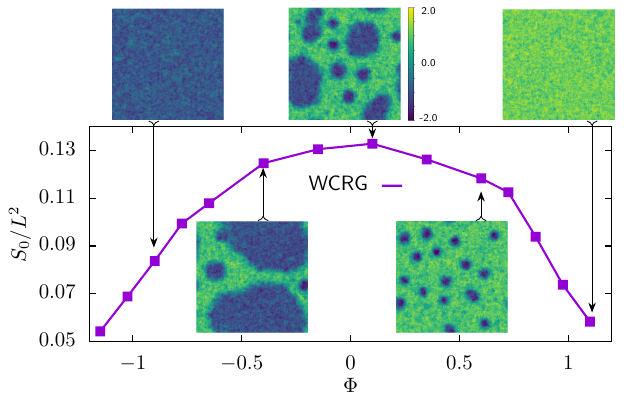}
\caption{The total entropy per site in the AMB+ for different mean fields $\Phi$ and $L=128.$ Snapshots show exemplary configurations at $\Phi=-0.9, -0.4, 0.1, 0.6$ and $1.1$ (same color code). }
\label{fig:active_total}
\end{figure}

{\it Active Model B+}:
We next study the Active Model B+ (AMB+), a paradigmatic field theory of nonequilibrium phase separation~\cite{tjhung2018cluster}. Compared with the equilibrium Model B \cite{hohenberg1977theory}, the AMB+ contains two additional terms in the flux which break detailed balance, as explained in Appendix \ref{app:activeBplus}. The AMB+ features a microphase separation with a finite characteristic domain size $\xi$ in the nonequilibrium steady state. Similar to the $\varphi^4$ model, the AMB+ is spatially discretized on a two-dimensional square lattice. The control parameter is the conserved mean field $\Phi = L^{-2}\sum_i \varphi_0(i)$, where $\varphi_0(i)$ is the local field at lattice site $i$. 
As $\Phi$ is increased, the system exhibits (i) a percolating-domain structure, (ii) bubbles with a broad distribution of sizes, and (iii) bubbles of approximately uniform size, as illustrated by the snapshots in Fig.~\ref{fig:active_total}. Importantly, the observed phase behavior is not symmetric under the transformation $\Phi \to -\Phi$.
We analyze this model within the WCRG framework, which allows us to reconstruct the steady-state probability distribution even when its analytical form is completely unknown, as demonstrated in Ref.~\cite{brossollet2025effective}. This enables us to compute multiscale and conditional entropies and thereby characterize the observed physical behavior.

In Fig.~\ref{fig:active_total}, we show the total entropy per site, $S_0/L^2$, estimated using WCRG. Its approximately symmetric shape is reminiscent of the binary entropy function: the entropy is maximal when the low- and high-$\Phi$ phases occupy comparable fractions of the system and decreases as either phase becomes dominant. However, the total entropy does not readily reveal the rich spatial ordering in this system, particularly the asymmetry of its phase behavior.

\begin{figure}
\includegraphics[width=\linewidth]{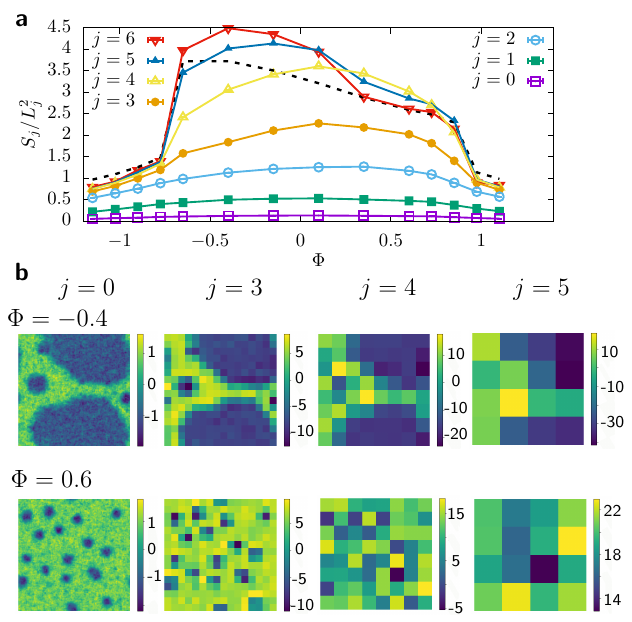}
\caption{(a) Multiscale entropy in the AMB+ compared to the correlation length $\xi$ for different mean fields $\Phi$ and $L=128.$ The black dotted line corresponds to $\ln \xi + \text{const.}$  (b) Snapshots show exemplary coarse-grained configurations at $\Phi=-0.4$ and $0.6$ for different scales $j$. }
\label{fig:active_multi}
\end{figure}

We therefore analyze and plot the multiscale entropies per site, $S_j/L_j^2$, in Fig.~\ref{fig:active_multi}(a). For $|\Phi|\simeq 1$, where either the low-$\Phi$ phase ($\Phi\simeq -1$) or the high-$\Phi$ phase ($\Phi\simeq 1$) dominates, the entropy per coarse-grained site, $S_j/L_j^2$, is small and nearly independent of $j$ for $j>2$. Since $L_{j+1}^2=L_j^2/4$, this implies $S_j\approx 4S_{j+1}$: the entropy at the finer scale is approximately four times that at the next coarser scale, in proportion to the number of sites. This behavior is consistent with weak spatial correlations and approximate statistical independence among coarse-grained sites at these scales. By contrast, an increase in the entropy per site, $S_{j+1}/L_{j+1}^2>S_j/L_j^2$, would suggest that spatial correlations and structure remain relevant at the scales considered.
For intermediate $\Phi$, the multiscale entropies reveal a pronounced asymmetry at coarser scales, $j>3$. Upon increasing $\Phi$ from the low-$\Phi$ homogeneous phase, the transition to the percolating-domain regime is accompanied by a sharp increase in $S_j/L_j^2$, which becomes more pronounced at larger $j$. This behavior is consistent with percolating structures that extend across a broad range of length scales and remain visible under successive coarse-graining steps (see $\Phi=-0.4$ in Fig.~\ref{fig:active_multi}(b)).

By contrast, upon approaching the coexistence region from high $\Phi$, bubbles with a relatively well-defined characteristic size $\xi$ emerge, and the increase in $S_j/L_j^2$ is more modest. Once the coarse-graining scale $2^j$ exceeds $\xi$, the entropy per site becomes nearly independent of $j$, with a weak nonmonotonic variation. This behavior is consistent with the disappearance of recognizable bubble patterns in the coarse-grained fields at scales larger than the characteristic bubble size (see $\Phi=0.6$ in Fig.~\ref{fig:active_multi}(b)).

To quantitatively compare the multiscale entropy with the characteristic domain size, we estimate $\xi$ from the decay of the two-point spatial correlation function and plot it in Fig.~\ref{fig:active_multi}(a). The variation of $\xi$ closely tracks that of the multiscale entropy at large scales, confirming that the scale-dependent entropy $S_j$ captures spatial information associated with nonequilibrium pattern formation.
In the Appendix~\ref{app:activeBplus}, we also plot the conditional entropy per site, $\overline{S}_j/(3L_j^2)$, which exhibits a similar trend. Thus, our multiscale-entropy framework reveals complex phase behavior even in systems far from equilibrium.

{\it Conclusion and discussion}:
We develop a multiscale entropy framework based on wavelet conditional renormalization group (WCRG), which provides a precise decomposition of the total entropy into scale-by-scale contributions. The resulting multiscale entropy serves as a generic scale-dependent order parameter, characterizing spatial ordering across scales in physical systems, including those far from equilibrium. Leveraging recent advances in generative machine learning, this framework enables accurate entropy estimation even when the underlying probability distribution is not known.

Future work could further elucidate the theoretical properties of the multiscale entropy and its connections to renormalization group theory, as well as extend the methodology to a broader range of nonequilibrium systems in active matter, biophysics, and complex systems more generally. As a proof of concept, Appendix~\ref{app:opinion_dynamics} presents results demonstrating how the multiscale entropy characterizes the phase behavior of an agent-based opinion dynamics model~\cite{perrier2024phase}.

In this work, we use a simple energy ansatz for WCRG modeling (see Appendix~\ref{sec:energy_based}), which is sufficient to compute the multiscale and conditional entropies of the physical systems studied here without neural networks. Extending the multiscale entropy analysis to more complex data, such as natural images, may require more expressive WCRG models based on richer energy ansatzes~\cite{lempereur2026hierarchic} or neural-network parameterizations~\cite{guth2022wavelet}.

\begin{acknowledgments}
We thank Giulio Biroli and Srikanth Sastry for very helpful discussions.
We thank the support by MIAI@Grenoble Alpes and the Agence Nationale de la Recherche under France 2030 with the reference ANR-23-IACL-0006 (MO). This research was partly funded by the Austrian Science Fund (FWF) 10.55776/PAT1139125 (GJ).
\end{acknowledgments}

\section*{Data Availability}

The source code used in this paper is openly available at \href{https://github.com/GerhardJung/WCRG_model}
{\url{https://github.com/GerhardJung/WCRG_model}}. 

\bibliography{ref}

\clearpage

\appendix

\section{Discrete wavelet transform}
\label{sec:DWT}

We briefly summarize the discrete wavelet transform in $d$ dimensions and its orthogonal decomposition property~\cite{mallat1999wavelet}.

We consider a field $\varphi_0 \in \mathbb{R}^{L^d}$ defined on a $d$-dimensional box of linear size $L$. We consider periodic boundary conditions.
In $d=2$, this field can be interpreted as an image. The scalar value at site $i$ is denoted by $\varphi_0(i)$.
Similarly, we denote by $\varphi_j \in \mathbb{R}^{L_j^d}$ the field defined at scale $j$ on a $d$-dimensional box of linear size $L_j = L/2^j$.
The scalar value of this field at site $i$ is denoted by $\varphi_j(i)$.

We wish to construct an orthogonal decomposition of $\varphi_{j-1} \in \mathbb{R}^{L_{j-1}^d}$ into coarse-grained and fine-scale contributions. To this end, we introduce an orthogonal projection operator $\mathcal{P}$ satisfying $\mathcal{P}^2=\mathcal{P}$ and $\mathcal{P}^T=\mathcal{P}$, where the superscript $T$ denotes transpose. We also introduce its complementary projector $\overline{\mathcal{P}}$, defined by $\mathcal{P}+\overline{\mathcal{P}}=I$,
where $I$ is the identity operator. With these operators, the field $\varphi_{j-1}$ can be decomposed as
\begin{equation}
\varphi_{j-1}
=
\mathcal{P}\varphi_{j-1}
+
\overline{\mathcal{P}}\varphi_{j-1}.
\end{equation}
The first term corresponds to the coarse-grained contribution, while the second term corresponds to the fine-scale contribution.

To construct such projectors, we introduce a low-pass filter $G$, which maps $\varphi_{j-1}$ to the coarse-grained field $\varphi_j \in \mathbb{R}^{L_j^d}$ as
\begin{equation}
\varphi_j = \gamma_j^{-1} G \varphi_{j-1},
\label{eq:WT_coarse}
\end{equation}
where $\gamma_j$ is a normalization factor. Similarly, we introduce a high-pass filter $\overline{G}$, which maps $\varphi_{j-1}$ to the wavelet field $\overline{\varphi}_j \in \mathbb{R}^{(2^d-1)L_j^d}$ as
\begin{equation}
\overline{\varphi}_j = \gamma_j^{-1} \overline{G} \varphi_{j-1}.
\label{eq:WT_fine}
\end{equation}

We then construct the projection operators as
\begin{equation}
\mathcal{P}
=
G^T G,
\qquad
\overline{\mathcal{P}}
=
\overline{G}^T \overline{G}.
\end{equation}
The filters $G$ and $\overline{G}$ are chosen such that $G^T G + \overline{G}^T \overline{G}=I$, 
together with the orthonormality conditions $GG^T = I$, $\overline{G}\overline{G}^T = I$, and $G\overline{G}^T=0$.
Here, the identity operators act on the corresponding coarse and fine spaces. In practice, we use the filters of the Daubechies D4 wavelet in this study~\cite{daubechies1992ten}.

With these definitions, the original field can be reconstructed as
\begin{equation}
\varphi_{j-1}
=
\gamma_j G^T \varphi_j
+
\gamma_j \overline{G}^T \overline{\varphi}_j .
\label{eq:IWT}
\end{equation}

Equations~(\ref{eq:WT_coarse}) and~(\ref{eq:WT_fine}) define the discrete wavelet transform, while Eq.~(\ref{eq:IWT}) defines the corresponding inverse discrete wavelet transform. Thus, the two representations, $\varphi_{j-1}$ and $(\varphi_j,\overline{\varphi}_j)$, are equivalent and can be regarded as a change of variables.

\vspace{0.5cm}
\section{Multiscale distributions}
\label{sec:multiscale}

We consider the multiscale probability distributions using a discrete wavelet transform~\cite{marchand2023multiscale}.
Since the two representations $\varphi_{j-1}$ and
$(\varphi_j,\overline{\varphi}_j)$ are equivalent, i.e., they are related by an invertible change of variables, the probability distribution of $\varphi_{j-1}$, denoted by
$p_{j-1}(\varphi_{j-1})$, can be expressed in terms of the conditional probability distribution of $\overline{\varphi}_j$ given $\varphi_j$, denoted by
$\overline{p}_j(\overline{\varphi}_j|\varphi_j)$, as
\begin{equation}
p_{j-1}(\varphi_{j-1})
=
\alpha_j
\overline{p}_j(\overline{\varphi}_j|\varphi_j)
p_j(\varphi_j),
\label{eq:conditional_distribution}
\end{equation}
where $\alpha_j=\gamma_j^{-L_{j-1}^d}$ is the normalization factor associated with the change of variables. With this convention, the integration measure transforms as
\begin{equation}
\int d\varphi_{j-1}(\cdots)
=
\alpha_j^{-1}
\int d\varphi_j
\int d\overline{\varphi}_j(\cdots).
\end{equation}
Thus, $\alpha_j^{-1}$ is the Jacobian factor appearing in the transformation from $\varphi_{j-1}$ to $(\varphi_j,\overline{\varphi}_j)$.

By iterating Eq.~\eqref{eq:conditional_distribution} from $j=1$ up to the largest scale $j=J$, we obtain
\begin{equation}
p_0(\varphi_0)
=
\alpha \, p_J(\varphi_J)
\prod_{j=1}^J
\overline{p}_j(\overline{\varphi}_j|\varphi_j) ,
\end{equation}
where $\alpha = \prod_{j=1}^J \alpha_j$.
This equation provides a natural factorization of the microscopic distribution $p_0(\varphi_0)$ into a product of conditional distributions across scales together with the coarsest scale distribution $p_J(\varphi_J)$.

\vspace{0.5cm}
\section{Energy-based modeling}
\label{sec:energy_based}

\subsection{Boltmann form}

To estimate the probability distributions across scales,
$p_j(\varphi_j)$ and
$\overline{p}_j(\overline{\varphi}_j|\varphi_j)$,
we employ energy-based modeling~\cite{lecun2006tutorial}, representing these distributions in the
Boltzmann form familiar from statistical mechanics.
This representation does not assume thermal equilibrium and also applies
to nonequilibrium stationary states: the effective energy parametrizes
the probability distribution and need not correspond to a physical energy.

For $p_j(\varphi_j)$, we write
\begin{equation}
    p_j(\varphi_j)
    =
    \frac{e^{-E_j(\varphi_j)}}{Z_j},
    \label{eq:energy_based_coarse}
\end{equation}
where $Z_j$ is the normalization factor, or partition function.
The function $E_j(\varphi_j)$ is the corresponding energy (or effective Hamiltonian),
whose functional form and parametrization will be discussed below.
We fix the additive constant of the energy by imposing
\begin{equation}
    E_j(\varphi_j)\big|_{\varphi_j=0}=0.
\end{equation}

Similarly, for the conditional distribution
$\overline{p}_j(\overline{\varphi}_j|\varphi_j)$, we write
\begin{equation}
    \overline{p}_j(\overline{\varphi}_j|\varphi_j)
    =
    \frac{
        e^{-\overline{E}_j(\varphi_{j-1})}
    }{
        \overline{Z}_j(\varphi_j)
    },
    \label{eq:conditional_distribution_appendix}
\end{equation}
where $\overline{E}_j(\varphi_{j-1})$ is the associated conditional
energy function, whose functional form will be discussed below together
with that of $E_j(\varphi_j)$. We fix its additive constant by imposing
\begin{equation}
    \overline{E}_j(\varphi_{j-1})
    \big|_{\overline{\varphi}_j=0}
    =
    \overline{E}_j(\mathcal{P}\varphi_{j-1})
    =
    0.
    \label{eq:conditional_energy_imposed}
\end{equation}
The normalization factor $\overline{Z}_j(\varphi_j)$ depends on the
coarse-grained field $\varphi_j$.

We express this dependence in terms
of a (conditional) free-energy function $\overline{F}_j(\varphi_j)$ as
\begin{equation}
    \overline{Z}_j(\varphi_j)
    =
    \overline{z}_j
    e^{-\overline{F}_j(\varphi_j)},
\end{equation}
where $\overline{z}_j$ is a constant and we choose
\begin{equation}
    \overline{F}_j(\varphi_j)\big|_{\varphi_j=0}=0.
\end{equation}
Therefore, the conditional distribution can be written as
\begin{equation}
    \overline{p}_j(\overline{\varphi}_j|\varphi_j)
    =
    \frac{
        e^{-\overline{E}_j(\varphi_{j-1})
        +\overline{F}_j(\varphi_j)}
    }{
        \overline{z}_j
    }.
    \label{eq:energy_based_conditional}
\end{equation}

We next establish a relation between $E_j(\varphi_j)$ and
$\overline{E}_j(\varphi_{j-1})$ using
Eq.~(\ref{eq:conditional_distribution}).
Substituting Eqs.~(\ref{eq:energy_based_coarse}) and
(\ref{eq:energy_based_conditional}) into
Eq.~(\ref{eq:conditional_distribution}), and using
Eq.~(\ref{eq:conditional_energy_imposed}), we obtain
\begin{equation}
   \overline{E}_j(\varphi_{j-1})
   =
   E_{j-1}(\varphi_{j-1})
   -
   E_{j-1}(\mathcal{P}\varphi_{j-1}) .
   \label{eq:connect_energies}
\end{equation}
Thus, the functional form, or parametrization, of $\overline{E}_j(\varphi_{j-1})$ is determined by that of $E_j(\varphi_j)$.

\subsection{Energy ansatz}

We next describe the parametrization of the energy functions, following Ref.~\cite{marchand2023multiscale}.
For $E_j(\varphi_j)$, we assume a quadratic interaction term and a nonlinear potential term:
\begin{equation}
    E_j(\varphi_j) = \frac{1}{2} \varphi_j^T K_j \varphi_j + C_j^T V(\varphi_j),
    \label{eq:energy_ansatz}
\end{equation}
where $K_j$ is the interaction matrix, $C_j$ is a vector of coefficients, and
$V(\varphi) = (V_1(\varphi), V_2(\varphi), \ldots, V_s(\varphi))$.
Thus, $C_j^T V(\varphi_j) = \sum_{n=1}^s C_{j,n} V_n(\varphi_j)$.
For compactness, we write
\begin{align}
    E_j(\varphi_j) &= \theta_j^T U_j(\varphi_j), \label{eq:E_compact}\\
    \theta_j &= (K_j/2, C_j), \\
    U_j(\varphi_j) &= (\varphi_j\varphi_j^T, V(\varphi_j)).
\end{align}
This compact form expresses the energy function as the inner product of the parameters $\theta_j$ to be determined and the basis functions $U_j(\varphi_j)$.

In this paper, we employ a local potential ansatz:
\begin{equation}
    V_n(\varphi_j) = \sum_{i=1}^{L_j^d} v_n(\varphi_j(i)).
\end{equation}
Specifically, we use hat functions $v_n(x) = h(x-na)$, each with support of width $2a$, where
\begin{equation}
    h(x) = \max\{x+a,0\} - 2\max\{x,0\} + \max\{x-a,0\}.
\end{equation}

We express $\overline{E}_j(\varphi_{j-1})$ in a compact form analogous to that of $E_j(\varphi_j)$:
\begin{equation}
    \overline{E}_j(\varphi_{j-1})
    = \overline{\theta}_j^T \overline{U}_j(\varphi_{j-1}).
    \label{eq:bar_E_compact}
\end{equation}
The parameters $\overline{\theta}_j$ and basis functions $\overline{U}_j$ are related to those of $E_{j-1}$ through Eq.~(\ref{eq:connect_energies}).
Substituting Eqs.~(\ref{eq:E_compact}) and~(\ref{eq:bar_E_compact}) into Eq.~(\ref{eq:connect_energies}), and defining
\begin{equation}
    \overline{U}_j(\varphi_{j-1})
    = U_{j-1}(\varphi_{j-1})
    - U_{j-1}(\mathcal{P}\varphi_{j-1}),
\end{equation}
we obtain
\begin{equation}
    \overline{\theta}_j^T \overline{U}_j(\varphi_{j-1})
    = \theta_{j-1}^T \overline{U}_j(\varphi_{j-1}),
    \label{eq:theta_from_bar_theta}
\end{equation}
which allows us to obtain $\theta_{j-1}$ from $\overline{\theta}_j$.

Using Eq.~(\ref{eq:energy_ansatz}), we obtain the explicit expression
\begin{align}
    \overline{\theta}_j^T \overline{U}_j(\varphi_{j-1})
    &= \frac{\gamma_j^2}{2}
    \overline{\varphi}_j^T \overline{G} K_{j-1}
    \overline{G}^T \overline{\varphi}_j  + \gamma_j^2 \varphi_j^T G K_{j-1}
    \overline{G}^T \overline{\varphi}_j \nonumber \\
    &\quad + C_{j-1}^T
    \left[V(\varphi_{j-1}) - V(\mathcal{P}\varphi_{j-1})\right].
    \label{eq:conditional_energy_parameterized}
\end{align}

\subsection{Parameter estimation}
\label{app:learning_WCRG}

The above energy parameterizations allow us to parameterize the corresponding probability distributions. For $p_j(\varphi_j)$, we introduce the model
\begin{equation}
    p_{\theta_j}(\varphi_j)
    = \frac{1}{Z_{\theta_j}}
    e^{-\theta_j^T U_j(\varphi_j)} .
    \label{eq:WCRG}
\end{equation}
The parameter $\theta_j$ can be estimated by maximizing the log-likelihood, or equivalently, by minimizing the Kullback--Leibler (KL) divergence,
$\min_{\theta_j} D_{\rm KL}(p_j \| p_{\theta_j})$, via gradient descent. We performed this only for $j=J$.

For the conditional distribution $\overline{p}_j(\overline{\varphi}_j|\varphi_j)$, we introduce the model
\begin{equation}
    \overline{p}_{\overline{\theta}_j}(\overline{\varphi}_j|\varphi_j)
    = \frac{
        e^{-\overline{\theta}_j^T \overline{U}_j(\varphi_{j-1})}
    }{
        \overline{Z}_{\overline{\theta}_j}(\varphi_j)
    } .
\end{equation}
Together with the marginal distribution $p_j(\varphi_j)$, this defines a model for $p_{j-1}(\varphi_{j-1})$,
\begin{equation}
    p_{\overline{\theta}_j}(\varphi_{j-1})
    = \overline{p}_{\overline{\theta}_j}(\overline{\varphi}_j|\varphi_j) \,
    p_j(\varphi_j) .
\end{equation}
The parameter $\overline{\theta}_j$ is then estimated by minimizing the KL divergence,
$\min_{\overline{\theta}_j} D_{\rm KL}(p_{j-1} \| p_{\overline{\theta}_j})$.
In practice, we first obtain an initial estimate of $\overline{\theta}_j$ using the score-matching scheme of Ref.~\cite{brossollet2025effective}. We then refine this estimate by minimizing $D_{\rm KL}(p_{j-1} \| p_{\overline{\theta}_j})$ using gradient descent, with the required expectations evaluated by Markov Chain Monte Carlo (MCMC) sampling of the wavelet field $\overline{\varphi}_j$. Since the energy function in Eq.~(\ref{eq:bar_E_compact}) is linear in $\overline{\theta}_j$, minimizing the KL divergence is a convex optimization problem. Combined with efficient MCMC sampling of the wavelet field, this allows us to estimate the parameters accurately.

We introduce an algorithmic improvement over Ref.~\cite{marchand2023multiscale} for MCMC sampling of the wavelet fields.
An important advantage of the Daubechies D4 wavelets \cite{daubechies1992ten} used in this work is their localization in both real and Fourier space. This motivates introducing a spatial cutoff $r_c$ in the Gaussian term of Eq.~(\ref{eq:conditional_energy_parameterized}), beyond which we set the interactions to zero. For a fixed cutoff, the computational cost of evaluating the energy difference for a local MCMC update is independent of the system size, analogous to local updates in the nearest-neighbor Ising model.
With this cutoff, we can learn models with $L=128$ within a single day on a single CPU processor. By comparison, the previous implementation presented in Ref.~\cite{marchand2023multiscale} required several weeks on 40 CPU processors for $L=64$. This improvement should make it feasible to learn models with $L=512$ and beyond using moderate computational resources, opening up possibilities for future applications. We use $r_c=2$ for the $\varphi^4$ model, $r_c=3$ for AMB+ and the HK opinion dynamics model.

The source code used in this paper is openly available at
\href{https://github.com/GerhardJung/WCRG_model}
{\url{https://github.com/GerhardJung/WCRG_model}}.

\section{Multiscale and conditional entropies}
\label{sec:multiscale_entropy}

We first provide the key equations defining the multiscale and conditional entropies. We then describe the numerical procedure used to compute them.

\subsection{Definitions}

We define the Shannon entropy of the coarse-grained field $\varphi_j$ at scale $j$, which we call the {\it multiscale entropy}, as
\begin{equation}
S_{j}
=
-\int d\varphi_{j}
p_{j}(\varphi_{j})
\ln p_{j}(\varphi_{j}) + \ln a_j  .
\label{eq:def_Shannon_coarse}
\end{equation}
Here, the constant $a_j = \tilde{\gamma}_j^{L_j^d}$, with $\tilde{\gamma}_j = \prod_{j'=0}^{j} \gamma_{j'}$, ensures that the entropy is independent of the normalization factors $\gamma_{j}$. Consequently, $S_j$ correctly represents the entropy of the underlying probability distribution.
Inserting Eq.~\eqref{eq:conditional_distribution} into Eq.~\eqref{eq:def_Shannon_coarse} for scale $j-1$, we obtain
\begin{equation}
S_{j-1}
=
S_j
+
\overline{S}_j ,
\label{eq:entropy_recurrence}
\end{equation}
where $\overline{S}_j$ denotes the {\it wavelet conditional entropy} at scale $j$, defined as
\begin{eqnarray}
\overline{S}_j
=
-\int d\varphi_j
p_j(\varphi_j)
\int d\overline{\varphi}_j
\overline{p}_j(\overline{\varphi}_j|\varphi_j)
\ln \overline{p}_j(\overline{\varphi}_j|\varphi_j) + \ln b_j, \label{eq:bar_S_appendix} \nonumber \\
\end{eqnarray}
where $\ln b_j = (2^d-1) \ln a_j$ ensures that $\overline{S}_j$ is also independent of the normalization factors.

By iterating Eq.~\eqref{eq:entropy_recurrence} from $j=1$ to $j=J$, we obtain
\begin{equation}
    S_0 = S_J + \sum_{j=1}^J \overline{S}_j ,
\end{equation}
which decomposes the total entropy $S_0$ into the entropy $S_J$ at the coarsest scale and the conditional entropy contributions $\overline{S}_j$ at each scale $j$.

\subsection{Numerical estimation of entropies from WCRG models}
\label{app:extracting_entropy}

At the coarsest scale, we estimate $S_J$ using the learned model $p_{\theta_J}(\varphi_J)$ as
\begin{equation}
    S_J \approx - \int d\varphi_J\,
    p_{\theta_J}(\varphi_J) \ln p_{\theta_J}(\varphi_J) + \ln a_J,
\end{equation}
which we evaluate by one-dimensional numerical integration over $\varphi_J \in \mathbb{R}$.

To compute $\overline{S}_j$, we first substitute Eq.~(\ref{eq:conditional_distribution_appendix}) into Eq.~(\ref{eq:bar_S_appendix}) to obtain
\begin{equation}
    \overline{S}_j
    = \langle \overline{E}_j(\varphi_{j-1}) \rangle_{p_{j-1}}
    + \langle \ln \overline{Z}_j(\varphi_j) \rangle_{p_j}
    + \ln b_j ,
\end{equation}
where $\langle \cdots \rangle_{p_j}
= \int d\varphi_j\, p_j(\varphi_j)(\cdots)$.
Numerically, we estimate this quantity using the learned WCRG models as
\begin{equation}
    \overline{S}_j \approx
    \langle \overline{\theta}_j^T
    \overline{U}_j(\varphi_{j-1}) \rangle_{p_{j-1}}
    + \langle \ln \overline{Z}_{\overline{\theta}_j}(\varphi_j)
    \rangle_{p_j}
    + \ln b_j ,
    \label{eq:bar_S_numerical}
\end{equation}
where
\begin{equation}
    \overline{Z}_{\overline{\theta}_j}(\varphi_j)
    = \int d\overline{\varphi}_j\,
    e^{-\overline{\theta}_j^T \overline{U}_j(\varphi_{j-1})} .
\end{equation}
We evaluate $\ln \overline{Z}_{\overline{\theta}_j}(\varphi_j)$
by thermodynamic integration at fixed $\varphi_j$, interpolating
between the exactly solvable Gaussian part of
Eq.~(\ref{eq:conditional_energy_parameterized}) (the 1st and 2nd terms) and the full
conditional energy, including the nonlinear term (the 3rd term).
The expectation values in Eq.~(\ref{eq:bar_S_numerical})
are evaluated by averaging over the corresponding fields
in the training dataset.

The multiscale entropies $S_j$ are then obtained recursively
from Eq.~(\ref{eq:entropy_recurrence}), using the conditional
entropies $\overline{S}_j$ and the coarsest-scale entropy $S_J$.

\section{$\varphi^4$ model}
\label{app:phi4_model}

We investigate the $\varphi^4$ model, defined by scalar fields $\varphi_0(i)$ on a two-dimensional lattice~\cite{milchev1986finite,troster2005free,kaupuvzs2016corrections}. The microscopic energy is given by
\begin{equation}
E_0(\varphi_0) = - \beta \sum_{\langle i,j\rangle} \varphi_0(i) \varphi_0(j) + \sum_{i=1}^{L^2} \left[ \varphi_0^2(i) + \lambda^\prime \left( \varphi_0^2(i) - 1 \right)^2 \right],
\end{equation}
where $\langle i,j\rangle$ denotes nearest-neighbor pairs. Throughout this work, we set $\lambda^\prime = 1$ and sample microscopic configurations using Monte Carlo simulations. The model exhibits a second-order phase transition at $\beta_c \simeq 0.67$. To calculate its total entropy $S_0$, we perform thermodynamic integration by continuously interpolating between $\lambda^\prime = 0$, corresponding to an exactly solvable Gaussian model, and $\lambda^\prime = 1$, corresponding to the original $\varphi^4$ model.

We vary $\beta \in [0.3,0.86]$, which plays a role analogous to inverse temperature, and consider system sizes $L=4,8,16,32,64,128$. For $\beta > \beta_c$, we train the WCRG model using only configurations with positive total magnetization, $\sum_i \varphi_0(i) > 0$, and subsequently add $\ln 2$ to the entropy to account for the two symmetry-related sectors.
This factor is accounted for in the total entropy $S_0$ and the multiscale entropies $S_j$, but cancels out in the conditional entropy $\overline{S}_j$ through Eq.~(\ref{eq:entropy_recurrence}).

\section{Active model B+ }
\label{app:activeBplus}

We study Active Model B+ (AMB+), an active field theory with a conserved scalar field introduced in Ref.~\cite{tjhung2018cluster}. We follow the discretization scheme and use the source code provided in Ref.~\cite{Tjhung2020Discretization}. As in the $\varphi^4$ model, the field is defined on a two-dimensional lattice. The dynamics of AMB+ includes explicit nonequilibrium contributions to the flux between neighboring lattice sites that break time-reversal symmetry. The strengths of these contributions are controlled by the parameters $\zeta$ and $\lambda$. For $\zeta=\lambda=0$, AMB+ reduces to the standard equilibrium Model B~\cite{hohenberg1977theory}. We choose the conserved spatially averaged field, $\Phi = L^{-2}\sum_i \varphi_0(i)$, as the control parameter, where $L^2$ is the number of lattice sites. (This quantity is denoted by $\phi_0$ in Ref.~\cite{tjhung2018cluster}.) We set $\zeta=4$ and $\lambda=1$, corresponding to the bottom panels of Fig.~5(a) in Ref.~\cite{tjhung2018cluster}, and use the same values for all other parameters as in that reference. We perform numerical simulations using the code provided in Ref.~\cite{Tjhung2020Discretization} and use the resulting steady-state configurations as the training dataset for WCRG to estimate the multiscale and conditional entropies.
\begin{figure}
\includegraphics[width=\linewidth]{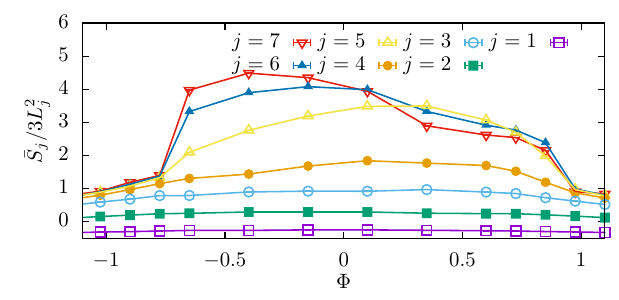}
\caption{Conditional wavelet entropies $\overline{S}_j$ extracted for the AMB+ using the WCRG model at different scales $j$. }
\label{fig:active_conditional}
\end{figure}

In Figs.~\ref{fig:active_total} and \ref{fig:active_multi} of the main text, we presented the total entropy $S_0$ and multiscale entropies $S_j$ for AMB+ with $L=128$. In Fig.~\ref{fig:active_conditional}, we show the conditional entropies $\overline{S}_j$, which exhibit trends similar to those of the multiscale entropies. In particular, they display a pronounced jump at the phase-separation transition at $\Phi \approx -0.65$, together with a noticeable dependence on the scale $j$ in the regime with percolating domains.

\section{Opinion dynamics model}
\label{app:opinion_dynamics}

We investigate a nonequilibrium model from a very different context: opinion dynamics, where each social agent $i$ holds an opinion $x_i$ that evolves through interactions with other agents, potentially leading to consensus \cite{starnini2026opinion}. The heterogeneous Hegselmann--Krause (HK) model is a bounded-confidence model in which agents on a lattice update their opinions by taking into account only neighbors whose opinions differ from their own by no more than their confidence threshold~\cite{Rainer2002-RAIODA,perrier2024phase}. Each agent has a confidence threshold $\epsilon_i$ that varies across the lattice and remains fixed in time, thus introducing quenched disorder. In Ref.~\cite{perrier2024phase}, this model was studied extensively, and its phase behavior was characterized using an order parameter defined as the average fraction of agents belonging to the largest cluster with equal opinions.

The model is defined on a square lattice with interactions extending up to third-nearest neighbors. Each agent $i$ has an opinion $x_i \in [-0.5,0.5]$, initially drawn from a uniform distribution. Each agent is also assigned a confidence threshold $\epsilon_i$, drawn randomly from the interval $[\epsilon_l,\epsilon_u]$ and held fixed throughout the simulation, thus introducing quenched disorder. The lower and upper bounds, $\epsilon_l$ and $\epsilon_u$, serve as control parameters governing the phase behavior.

Starting from a random initial state, each agent interacts with neighbors whose opinions lie within its confidence threshold. The effective neighborhood of agent $i$ at time $t$, including the agent itself, is defined as
\begin{equation}
I_i(t) = \{i\} \cup
\{j \in \mathcal{N}(i) \, | \, \, |x_i(t)-x_j(t)|<\epsilon_i\}.
\end{equation}
All agents update their opinions simultaneously according to
\begin{equation}
x_i(t+1) = \frac{1}{|I_i(t)|}
\sum_{j \in I_i(t)} x_j(t).
\end{equation}
Each agent thus adopts the average opinion of its effective neighborhood. Each simulation is run for $10^5$ time steps, by which time the system is assumed to have reached a steady state. The ensemble is constructed from configurations obtained after $10^5$ steps in independent runs with different random initial conditions. We fix $\epsilon_l=0.03$ and vary $\epsilon_u \in [0.03,0.55]$, corresponding to the red line in Fig.~1(b) of Ref.~\cite{perrier2024phase}. The system size is $L=128$.

\begin{figure}
\includegraphics[width=\linewidth]{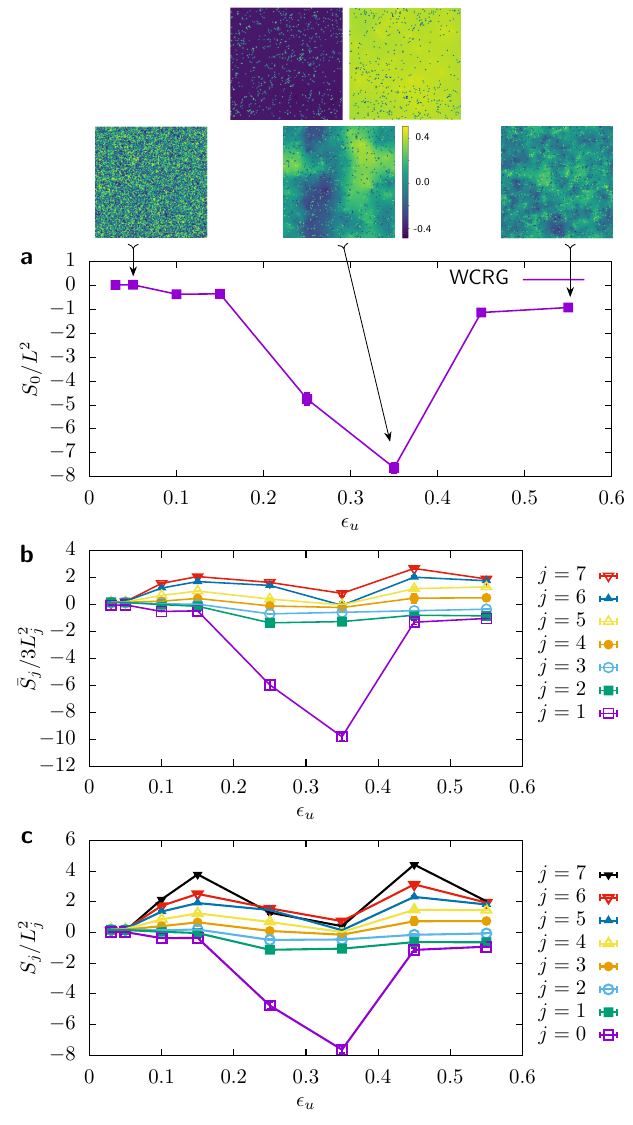}
\caption{Entropies per site in the heterogeneous Hegselmann--Krause opinion dynamics model as a function of the upper confidence bound $\epsilon_u$ for $L=128$.
(a) Estimated total entropy $S_0$. Snapshots show representative configurations at $\epsilon_u=0.05$, $0.35$, and $0.55$, using the same color scale. At $\epsilon_u=0.35$, three distinct phases coexist, as described in Ref.~\cite{perrier2024phase}.
(b) Conditional entropies $\overline{S}_j$ extracted from the WCRG model at different scales $j$.
(c) Multiscale entropies $S_j$ for the same system as in (b). }
\label{fig:opinion_entropy}
\end{figure}

At fixed $\epsilon_l=0.03$, increasing $\epsilon_u$ leads to a nonmonotonic evolution of consensus: the fragmented state at low confidence gives way to larger opinion clusters, whereas further increasing $\epsilon_u$ can weaken consensus. At intermediate values, such as $\epsilon_u=0.35$, different realizations can reach a broad, approximately symmetric opinion distribution, a skewed state dominated by an extremist opinion, or a state reached through a ``U-turn'' in which the dominant opinion reverses during the transient~\cite{perrier2024phase}.

Consistent with this picture, the total entropy in Fig.~\ref{fig:opinion_entropy}(a) reveals pronounced changes in the final opinion configurations as $\epsilon_u$ varies. For small $\epsilon_u$, agents largely retain their initial opinions because their confidence thresholds are typically much smaller than the opinion differences between neighbors. At large $\epsilon_u$, agents with small confidence thresholds and extreme opinions, $x_i \approx \pm 0.5$, can sustain opinion differences across the system and prevent global consensus. At intermediate $\epsilon_u$, global consensus can emerge in some realizations, as illustrated by the snapshots in Fig.~\ref{fig:opinion_entropy}(a), although other outcomes coexist. Consequently, the entropy is high at small $\epsilon_u$, decreases in the intermediate region where consensus is favored, and increases again at large $\epsilon_u$.

The conditional and multiscale entropies characterize the spatial structure of the opinion configurations, as shown in Fig.~\ref{fig:opinion_entropy}(b,c). For small $\epsilon_u$, the multiscale entropies $S_j$ collapse across scales, suggesting negligible spatial correlations. At large $\epsilon_u$, this collapse occurs only at coarse scales, $j>4$, suggesting a correlation length of approximately $\xi \approx 32$, consistent with the clustering visible in the snapshot at $\epsilon_u=0.55$. The boundaries between these more disordered regimes and the intermediate regime favoring consensus are marked by pronounced peaks in $S_j$ near $\epsilon_u \approx 0.15$ and $\epsilon_u \approx  0.45$. Similar to the behavior observed in the $\varphi^4$ model, these peaks suggest that multiscale entropies provide useful indicators for locating transitions between distinct phases.

This analysis demonstrates that multiscale entropies can characterize the phase behavior of an opinion dynamics model through a purely data-driven approach, without requiring the \emph{a posteriori} construction and measurement of complex order parameters.

\end{document}